\documentclass[12pt,preprint]{aastex}

\shorttitle{SiO maser and IR radiation }
\shortauthors{B.W.~Jiang}

\begin{document}

\title{Relation of SiO maser emission to IR radiation in evolved stars
based on the MSX observation}

\author{B.W. Jiang} \affil{National Astronomical Observatories,
Chinese Academy of Sciences, Beijing 100012, P.R.China}
\email{jiang@bao.ac.cn}

\begin{abstract}
Based on the space MSX observation in bands A(8$\mu$m), C(12$\mu$m),
D(15$\mu$m) and E(21$\mu$m), and the ground SiO maser observation of
evolved stars by the Nobeyama 45-m telescope in the v=1 and v=2 J=1-0
transitions, the relation between SiO maser emission and mid-IR
continuum radiation is analyzed. The relation between SiO maser
emission and the IR radiation in the MSX bands A, C, D and E is all
clearly correlated. The SiO maser emission can be explained by
a radiative pumping mechanism according to its correlation with 
infrared radiation in the MSX band A.
\end{abstract}

\keywords{infrared: stars --- masers --- stars: late-type}

\section{Introduction}

The relation between SiO maser emission and IR radiation was predicted
by the radiative pumping model for the SiO maser by \citet{deg76} in
which SiO maser is pumped by 
the $\triangle v=1$ vibrational transition at 8$\mu$m.  \citet{buj81}
analyzed a radiative pumping model in which the SiO molecules are 
located in the
inner circumstallar envelope and proposed that the inversion is achieved by
a direct population transfer from the $v-1$ state via the absorption
of stellar 8$\mu$m photons. This is followed by an optically thick
radiative decay back to $v-1$. Their calculation predict  
that the SiO maser peak intensity be smaller than the 8$\mu$m flux.

\citet{buj81} were also the first to look at whether there exists a relation between the SiO maser emission and
stellar IR radiation. Their analysis of
data from about 20 objects showed that the observations satisfy the
requirement to pump SiO maser radiatively, i.e. they lie below the
line S$_{\rm peak}${\footnotesize (SiO)} $\leq$ S{\footnotesize
(8$\mu$m)} \citep{buj81}.  \citet{buj87} assembled observational data
of more evolved stars and analyzed the relation between SiO maser
emission and radiation in some mid-infrared bands and found a
good correlation exists between them. However, analysis of the
integrated intensity of the bulge SiO maser sources resulted in large
scattering between the SiO maser integrated intensity and the infrared
fluxes at the IRAS bands 12, 25 and 60$\mu$m \citep{jia95}.

The progress in infrared space astronomy and SiO maser surveys makes it
possible now to revisit the relation between SiO maser and mid-infrared 
emission. The Midcourse Space Experiment (MSX) mission surveyed the
entire Galactic plane within $|b| \leq 4.5\degr$ in five infrared bands
B, A, C, D, and E, at 4, 8, 12, 15 and 21$\mu$m respectively
\citep{pri01}. In addition to its 30 times better spatial resolution
than IRAS, the most sensitive band A of MSX is centered at 8$\mu$m
where the proposed infrared radiation pumps the SiO J=1-0
maser. Furthermore, large scale searches for SiO maser emission 
from evolved stars in different parts of the Galactic plane 
have been carried out in the
J=1-0 rotational transition at the first and second
vibrationally excited states  by  the
Nobeyama 45-m telescope \citep{ita01}. This effort detected a large 
number of faint SiO
maser objects with F$_{12}\sim$ 1Jy (where F$_{\lambda}$ means the
flux intensity in Jy at band $\lambda$ and $\lambda$ refers to the
IRAS band 12$\mu$m, 25$\mu$m, MSX band A, B, C, D, E).

\section{Data}

All the SiO maser data in present study are based on the 
observation by using the
Nobeyama 45-m telescope. The combination of a large antenna and
sensitive detector at 43GHz in this system made it
feasible to detect an SiO maser emission of the J=1-0 rotational
transition as weak as 0.2~K. Since 1992 \citep{nak93}, this system has
been used to observe thousand of the IRAS PSC sources that are
candidates for evolved stars defined by their IRAS colors, and has
detected about 500 new SiO maser objects during the surveys of the
bulge \citep{izu94,izu95a,izu95b}, the inner disk
\citep{izu99,deg00a,deg00b}, the outer disk \citep{jia96, jia99}, and
the northern pole \citep{ita01}.

The SiO maser data in this study is compiled from
\citet{izu94,izu95a,izu95b,izu99,deg00a,jia96,jia99}, which 
mostly consists of sources in the Galactic plane. These studies 
achieved approximately
the same detection limit, with sensitivity of 
about 0.15~K at a 5$\sigma$ level that varies slightly
depending on the system condition, weather etc.  To guarantee the
quality of the data, tentative detections listed in these papers were
rejected and the final sample of SiO maser stars consists of 443
objects. Maser emission at both v=1 and v=2 J=1-0
transitions were detected in 421 of these stars while there are 11 stars
in which the SiO maser emission is detected at only the v=1 J=1-0
transition and another 11 stars at only v=2 J=1-0 transition. The
antenna efficiency is taken to be 50\%, which results in a conversion
factor $\sim$ 3.6Jy/K from antenna temperature to SiO maser flux intensity.

The infrared data are taken from the initial MSX Infrared Point Source
(Version 1.2) Catalog \citep{ega99}. This catalog provides 
photometric measurements in five infrared spectral bands to many SiO maser
stars, in particular most of the maser stars in the Galactic plane.
It was released in 1999 and is available through IPAC at
http://www.ipac.caltech.edu/ipac/msx/msx.html. This catalog contains
infrared observations on more than 300,000 sources, and lists the
flux density at the MSX bands with the quality index, variability
index etc. The MSX PSC1.2 appears to be complete above 0.2Jy and
about 50\% complete at a flux of about 0.17Jy, in the most sensitive
band A in the inner Galactic plane \citep{ega99}.  Details on the MSX
survey are referred to \citet{ega99}.

The SiO maser stars are cross-associated with the MSX PSC1.2 sources by
position coincidence. The position of SiO maser stars is adopted from
IRAS PSC since the HPBW of the 45-m telescope at 43GHz is about
40$\arcsec$, much bigger than a typical IRAS PSC error ellipse of
about 10$\times$20$\arcsec$ in radius. The position of the MSX PSC1.2 is
more accurate than the IRAS PSC. The MSX position uncertainty is on the order 
of 2.0$\arcsec$ in both the in-scan and cross-scan directions and 
the typical rms position error
cited in the MSX PSC1.2  is
about 2-4$\arcsec$.  The radius to search for the MSX
counterparts of the SiO maser stars is set to be 30$\arcsec$ and the
same for all the SiO maser stars for simplicity.

\begin{figure} 
\plotone{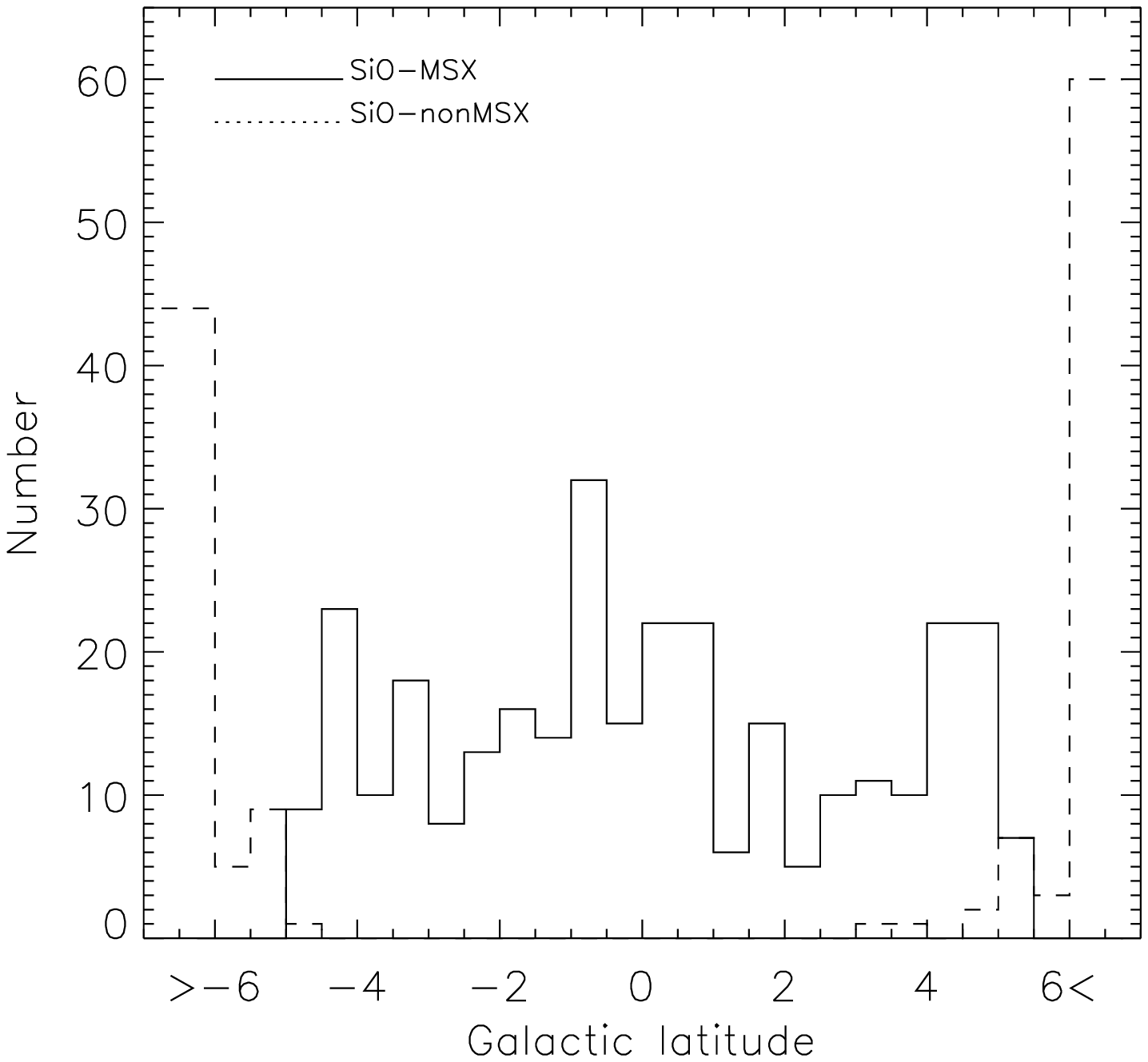} 
\caption{Distribution along the Galactic latitude of
the SiO stars with (full line) and without (dash line) MSX
association.} \label{fig1}  
\end{figure}

Within the search radius of 30$\arcsec$, 310 of the 443 SiO maser
stars are cross-associated and 133 are not cross-associated with the
MSX PSC sources. The distribution of these two groups of objects in
the Galaxy depends strongly on the Galactic latitude. The associated
stars are located at $|b|<6\degr$. The non-associated stars mostly
are located at $|b|>4.5\degr$, i.e. 131 of the 133 non-associated
stars have $|b|>4.5\degr$ and  only 2 
objects with $|b|<4.5\degr$ where the MSX survey fully covered (see
Fig.\ref{fig1}). 
The two non-associations with $|b|<4.5\degr$ are actually bright with
the flux density at IRAS 12$\mu$m band larger than 4Jy and their IRAS
PSC position should be relatively accurate. 
In order to know whether they were detected by MSX, a further check of 
the co-added MSX images was made and showed that they were actually 
detected by MSX 
clearly in all the A, C, D and E bands but missed in the catalog. 
 Among the 310 associated SiO maser stars, 4 objects are found to have
2 counterparts each within the search radius. The counterpart whose
flux intensity in the MSX C band is close to that at the IRAS 12$\mu$m
is regarded to be the right one because these two bands are quite
analogous. The other 306 stars have unique counterpart within this
search radius. There are five SiO maser stars whose photometric results  
in band
C are unavailable in the MSX PSC catalog possibly because of 
the faintness since their F$_{12}<$3.0Jy. The co-added
MSX images neither show clear evidence of detection. 
The photometric results in the MSX band A which is the key band for this study
 are available in the MSX PSC1.2 catalog for  
all the associated objects except IRAS 03469+5833. 
Since the counterpart of IRAS 03469+5833 is bright
in the MSX C band with F$_{\rm C}=$5.5 Jy close to
its IRAS 12$\mu$m intensity F$_{12}=$7.3Jy which also proves the
right association, the detection in band A is expected.  Indeed, the object 
is clearly detected in the MSX image in band A. The missing of bright sources
in the MSX PSC V1.2 may be ascribed to the incompleteness of the catalog.   
Further discussion is confined to the objects that are actually in the MSX 
PSC1.2.

The cross-associated sample consists of 310 stars, from which 288 are
detected with SiO maser emission at both the v=1 and v=2 J=1-0 transitions,
11 detected at only v=1 J=1-0 transition and 11 at only v=2 J=1-0
transition, and from which 309 stars are measured in the MSX A band, 305
measured in the MSX C band. There are 298 v=1 J=1-0 SiO maser stars with
association in the MSX A band and 298 v=2 J=1-0 SiO maser stars with
association in the MSX A band.

\begin{figure} \plotone{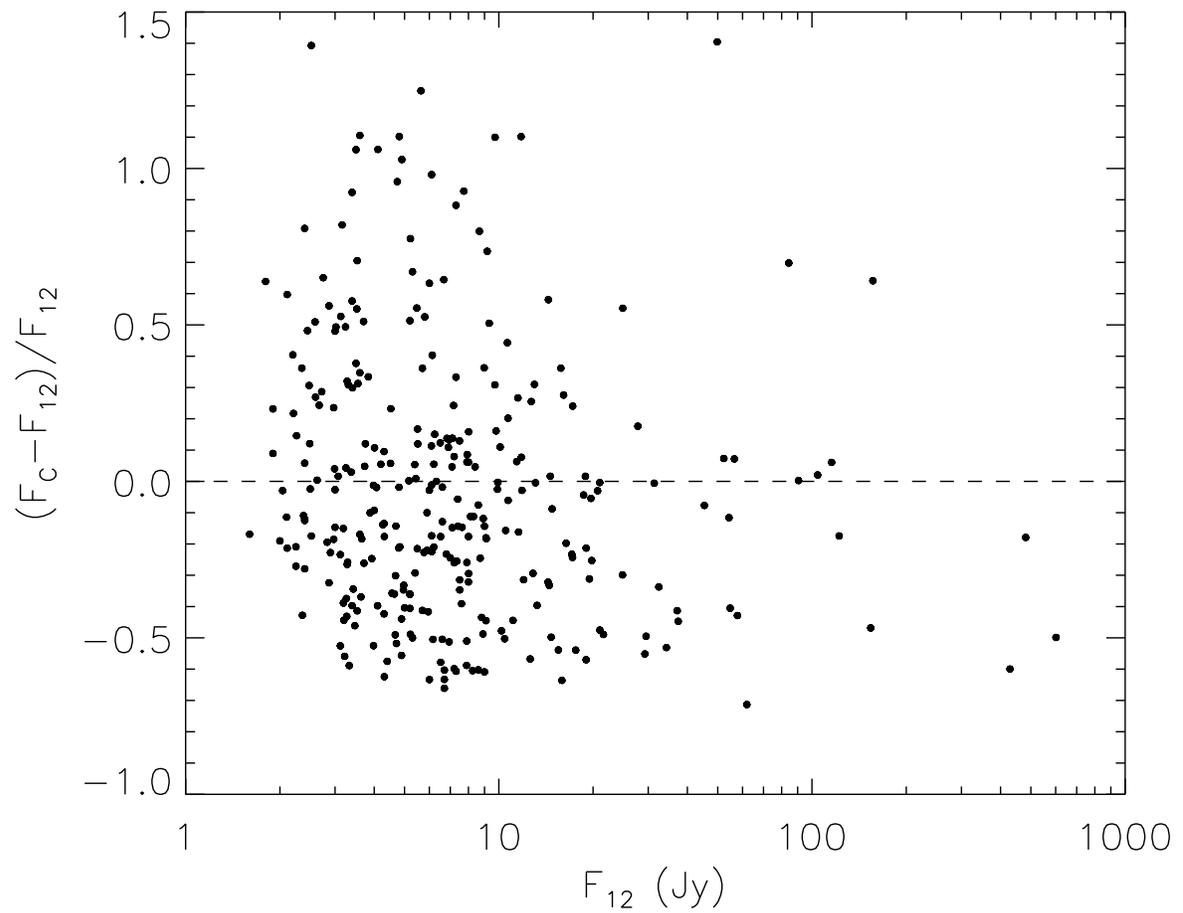} 
\caption{Relative difference of flux intensity in the
MSX C band to the IRAS 12$\mu$m band.} \label{fig2}
\end{figure}

Most of the sources are variables by comparing their flux intensity in
the MSX C band F$_{\rm C}$ with that at the IRAS 12$\mu$m
F$_{12}$. Though the MSX C band is not identical to the IRAS 12$\mu$m
band, it is designed to be a narrower analog of the IRAS 12$\mu$m
filter and its effective wavelength is just 12.13$\mu$m
\citep{pri01}.  Based on the MSX photometric accuracy, the flux differences
greater than 10\% is indicative of variability rather than photometric error
\citep{coh00}. The relative differences of F$_{\rm C}$ to F$_{12}$
mostly are greater than 10\% (Fig. \ref{fig2}), which means the differences between the MSX and IRAS
observations are intrinsically from the objects. Further supports to the sources being variable come from the independent redundant observations by MSX and IRAS to some of the objects. \citet{jia95} checked the
IRAS variability indexes of the SiO maser stars in the Galactic bulge by then
and found about two-thirds with the indexes greater than 90.  In the MSX PSC1.2 catalog, 167 of the 310 associated stars have the variability indexes being 1 in at least one band which means the variation over 3$\sigma$ \citep{ega99}.      Most, if not all, the
sources are variables.  The upper limit of ${\rm (F_C-F_{12})/F_{12}}$
is about 1.5 and this further implies that the amplitude of variation
at this band is not very large. The property of variation is
consistent with the SiO maser stars being mostly variable late-type
stars \citep{jia95}.

\section{Relation to infrared radiation}

\begin{figure} \plotone{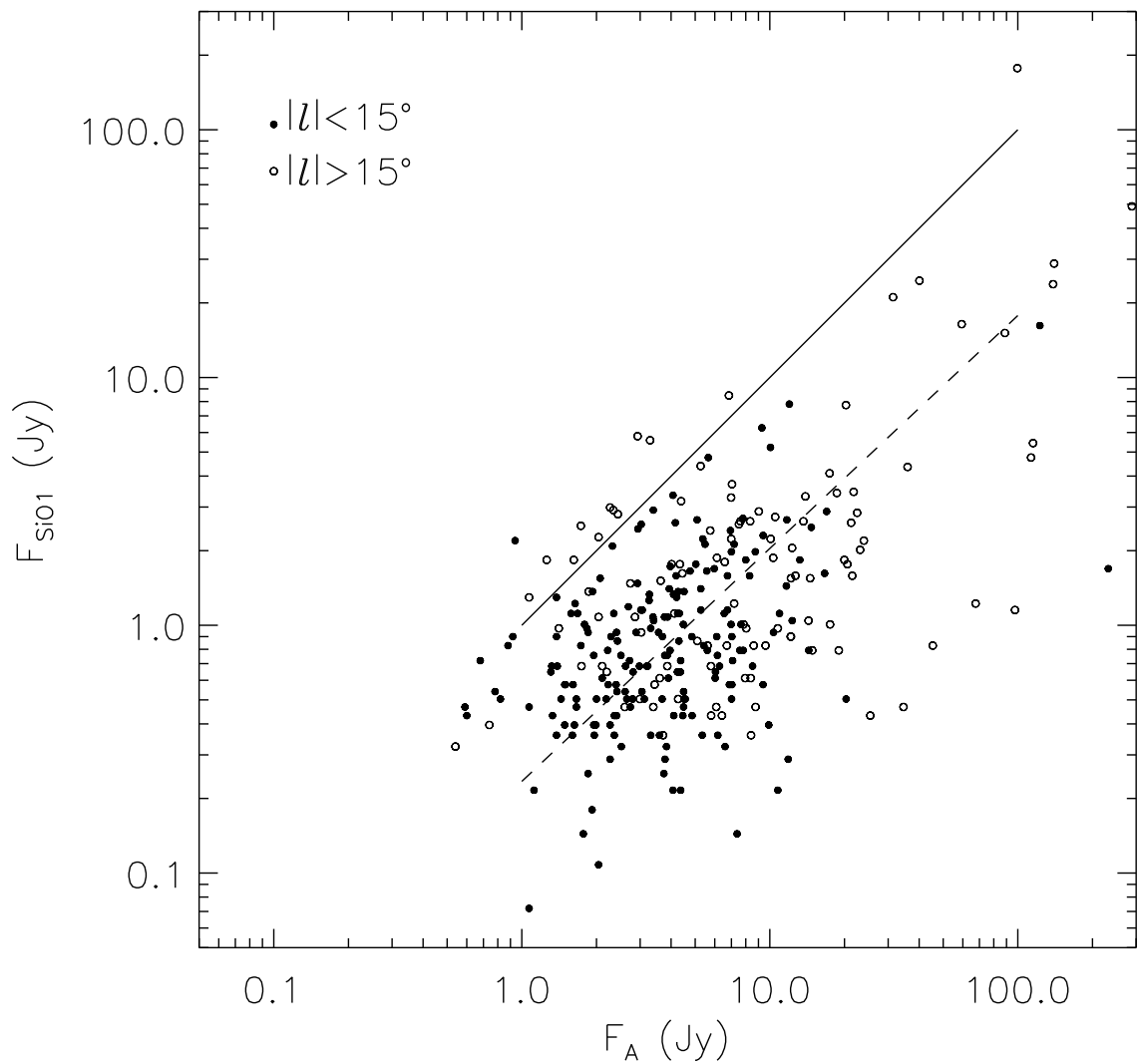}  
\caption{Relation of the v=1 J=1-0 SiO maser peak
intensity with the MSX A band flux is highly correlated. Dash line
shows the linear fitting result between ${\rm \log F_{SiO1}}$ and
${\rm \log F_A}$ by ordinary least-squares (OLS) bisector method and
the solid line borders the equal line. The objects with $|l|<15\degr$
and $|l|>15\degr$ are separated by signs small dot and open circle
respectively.}  \label{fig3}
\end{figure}

First of all, the relation of SiO maser peak intensity to the infrared
radiation in the MSX A band is discussed. Shown in Fig. \ref{fig3} are the
298 SiO maser stars that are detected the emission at the v=1 J=1-0
transition and the radiation in the MSX A band. The SiO maser
intensity ranges from about 0.1Jy to 100Jy, covering four orders of
magnitude. The linear correlation coefficient between F$_{\rm SiO1}$ (where
F$_{\rm SiO1 (2)}$ means the SiO maser peak intensity at the v=1 (2) J=1-0 
transition) 
and F$_{\rm A}$ is 0.44, which means they are clearly correlated from
the statistical point of view. A linear fitting relation is drawn by
the so-called ordinary least square (OLS) regression bisector method
\citep{iso90}, ${\rm \log F_{SiO1}=-0.63(\pm0.04)+0.94(\pm0.05)* \log
F_{A}}$.  The SiO maser peak intensity increases proportionally to the
infrared radiation. In addition, the SiO
maser intensity is systematically smaller than the infrared radiation at
8$\mu$m, as can also be seen from Fig.\ref{fig3} where the equal line is
solid, above the fitting dash line. \citet{buj81} calculated the
requirement for radiative pumping to work is ${\rm F_{SiO} \leq
F_{8\mu} }$. The relation between the SiO maser intensity and the MSX A
band flux supports the radiative pumping model to explain the
maser occurrence of the SiO v=1 J=1-0 transition. 
Although there are several
sources above the equal line in Fig.\ref{fig3}, the reason is 
possibly the variability of the sources because the MSX observation
didn't take place at the same phase of the variation as the SiO maser
observation. The variation of the objects and the different phases of
the SiO maser and MSX observations can also explain the scattering of
sources around the linear fitting line.

\begin{figure} \plotone{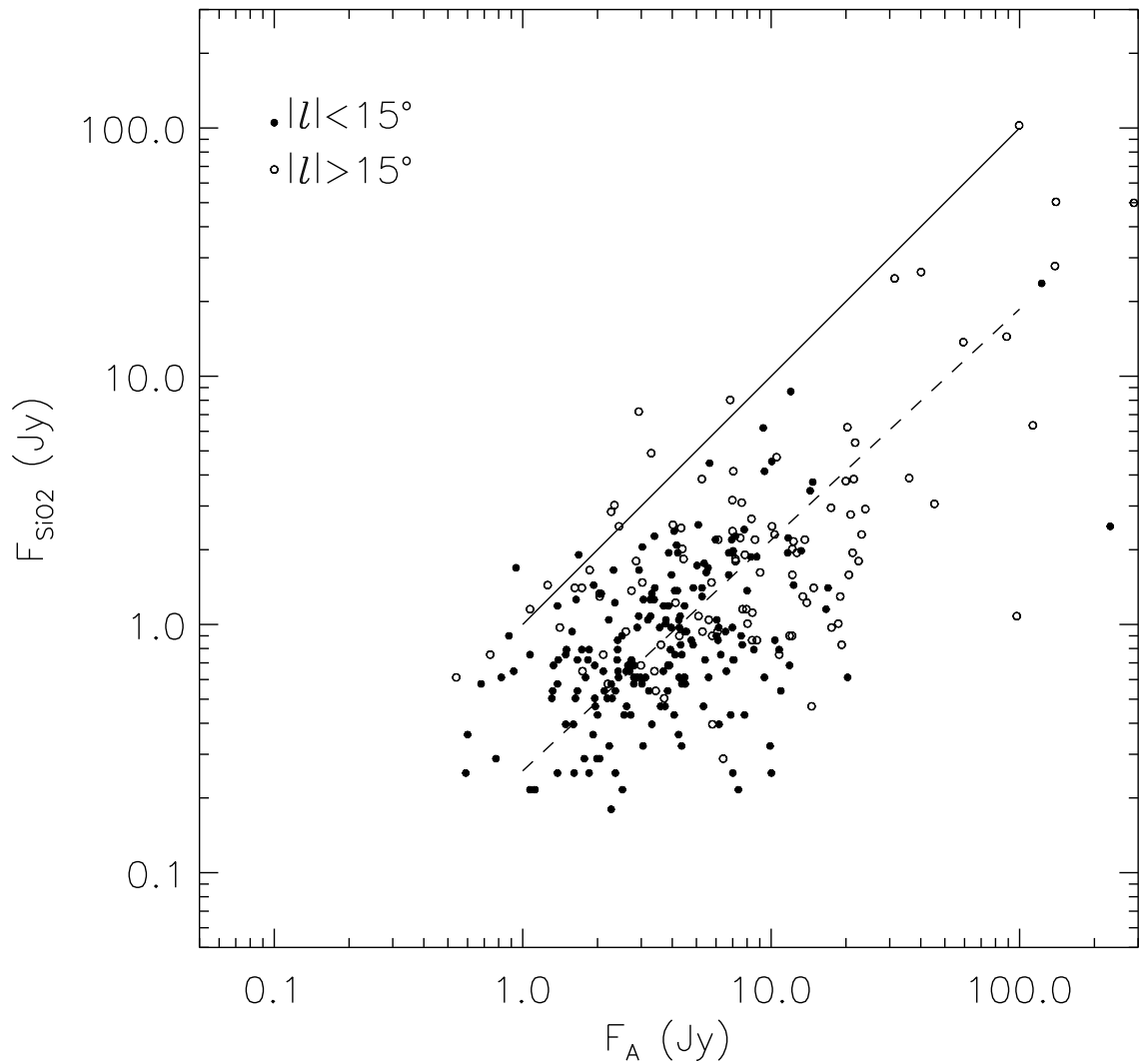}  
\caption{Relation of the v=2 J=1-0 SiO maser peak
intensity with the MSX A band flux is correlated as well as the v=1
J=1-0 maser. Convention of symbols is the same as in Fig.\ref{fig3}.}
\label{fig4} 
\end{figure}

A similar relation holds for the SiO maser peak intensity at the v=2
J=1-0 transition (Fig.\ref{fig4}). The linear correlation coefficient
between the maser intensity at this v=2 line F$_{\rm SiO2}$ and
F$_{\rm A}$ is 0.62, which again means a clearly correlated
relation. The OLS bisector linear regression method draws ${\rm \log
F_{SiO2}=-0.59(\pm0.03)+0.93(\pm0.04)* \log F_{A} }$.

In order to see if the relation remains the same in the disk and in
  the bulge, the sources with $|l|<15\degr$ which are supposed to be
  in the bulge are denoted by small dots and the sources with
  $|l|>15\degr$ which are supposed be in the disk are denoted by open
  circles in Fig.\ref{fig3}.  No apparent difference is found in the
  relation of the maser intensity to the infrared flux intensity in
  the MSX A band between these two groups of sources.

The relation of SiO maser intensity to the infrared radiation in other
MSX bands are examined except in the MSX band B. The MSX survey had
lower sensitivity in band B around 4$\mu$m \citep{ega99} which led to
the failure of detection to most of the SiO maser stars in this
band. In bands C, D and E, the relations between the SiO maser
emission and infrared radiation are all clearly correlated without
exception. The correlation coefficients of the SiO maser peak
intensity at the v=1 J=1-0 transition with the radiation intensity in
the MSX C, D and E bands are 0.44, 0.41 and 0.33 respectively, and the
corresponding coefficients at the v=2 line are 0.66, 0.64 and
0.56. All these correlation coefficients indicate tight correlation
from the statistical point of view.  The fact that the SiO maser
intensity is correlated with the radiation in any of the middle
infrared MSX bands can be expected if the radiation at these bands all
comes from the stellar continuum radiation other
than line emission. It seems no significant difference between the
relation to the intensity in the MSX A band and in other bands C, D,
E. This may mean that there is no necessity for the radiation through
$\triangle v=1$ transition at 8$\mu$m to be the pumping source
of the maser emission. As noted by \citet{buj87}, the general correlation 
of the SiO
maser intensity with the mid-infrared radiation may indicate the 
SiO maser is related to the amount of circumstellar matter in the evolved
stars.  Another phenomenon in the
relations is that the v=2 J=1-0 maser intensity generally exhibits
tighter correlation with smaller scattering with the infrared
radiation than the v=1 maser intensity and no model ever predicted
so. Whether this difference is true requires further investigations
such as simultaneous maser and infrared observations.

\section{Summary}

Based on new database in the mid-infrared provided by the MSX space
survey, the relations between the SiO maser peak intensity and
mid-infrared radiation in four MSX bands are analyzed. It is found
that the SiO maser intensity is correlated with mid-infrared
radiation, which confirms the conclusions from previous studies. The
result is consistent with the radiative pumping model for the SiO maser
emission in evolved stars.

\acknowledgments  The author acknowledges Alain Omont for
helpful discussion and the anonymous referee for useful suggestions. 
This work was principally supported by China's grant NKBRSF G19990754.

\clearpage

\end{document}